\documentclass[12pt]{article}
\usepackage{subfloat}
\pdfoutput = 1
\usepackage{graphics}
\usepackage{graphicx} 
\usepackage{amsmath}
\usepackage{amsfonts}   
\usepackage{amssymb}
\usepackage{multirow}
\usepackage{hhline}
\usepackage{color}
\usepackage{epsfig}
\usepackage{epstopdf}
\begin{document}
\date{Today}
\title{{\bf{\Large Effect of magnetic field on holographic insulator/superconductor phase transition in higher dimensional Gauss-Bonnet gravity }}}
\author{ {\bf {\normalsize Diganta Parai}$^{a}$
\thanks{digantaparai007@gmail.com}},\,
{\bf {\normalsize Debabrata Ghorai}$^{b}
$\thanks{debanuphy123@gmail.com, debabrataghorai@bose.res.in}},\,
{\bf {\normalsize Sunandan Gangopadhyay}$^{b}$
\thanks{sunandan.gangopadhyay@gmail.com, sunandan.gangopadhyay@bose.res.in}}\\
$^{a}$ {\normalsize Indian Institute of Science Education and Research Kolkata}\\{\normalsize Mohanpur, Nadia 741246, India}\\[0.2cm] 
$^{b}$ {\normalsize Department of Theoretical Sciences } \\{\normalsize S.N. Bose National Centre for Basic Sciences }\\{\normalsize JD Block, 
Sector III, Salt Lake, Kolkata 700106, India}\\[0.2cm]
}
\date{}

\maketitle
\begin{abstract}
{\noindent In this paper, we have investigated the effect of magnetic field numerically as well as analytically for holographic insulator/superconductor phase transition in higher dimensional Gauss-Bonnet gravity. First we have analysed the critical phenomena with magnetic field using two different numerical methods, namely, quasinormal modes method and the shooting method. Then we have carried out our calculation analytically using the St$\ddot{u}$rm-Liouville eigenvalue method. The methods show that marginally stable modes emerge at critical values of the chemical potential and the magnetic field satisfying the relation $\Lambda^2\equiv\mu^2-B$. We observe that the value of the chemical potential and hence the value of $\Lambda$ increases with higher values of the Gauss-Bonnet parameter and dimension of spacetime for a fixed mass of the scalar field. This clearly indicates that the phase transition from insulator to superconductor becomes difficult in the presence of the magnetic field for higher values of the Gauss-Bonnet parameter and dimension of spacetime. Our analytic results are in very good agreement with our numerical results.}
\end{abstract}
\vskip 1cm

\section{Introduction}
An enormous amount of investigation has been carried out on the AdS/CFT correspondence \cite{adscft1}-\cite{ssg} and applications of this duality in condensed matter physics to understand the basic properties of high $T_c$ superconductors \cite{adscft5}-\cite{nrt}. It is a map which relates strongly coupled systems to weakly coupled systems. This theoretical insight has been exploited to explain phase transition in strongly coupled system.  
The important result that one gets by using the gauge/gravity correspondence is the formation of a condensate below a certain temperature called the critical temperature. The idea is to construct a gravity theory in one higher dimension and study its properties. The duality is then applied to extract the properties of the boundary theory. \\
\noindent There has been a lot of work on the holographic metal/superconductor transition. However to describe an insulator/superconductor phase transition one has to consider a holographic model in the bulk AdS soliton background \cite{cai-so}-\cite{cai1m}. Further in \cite{rlhy}, the response of magnetic field on this phase transition has been studied in Einstein gravity background. However, we note that the effect of magnetic field in Gauss-Bonnet (GB) gravity in arbitrary spacetime dimensions would be important to look at. The reason for this is that GB gravity is a higher curvature gravity theory in higher spacetime dimensions and the Mermin-Wagner theorem claims that the phase transition is affected by higher curvature corrections \cite{hs8}. Study incorporating the GB gravity background effects on holographic insulator/superconductor phase transition without magnetic field in higher dimensions has been done in \cite{dp2}. The main purpose of this investigation is to see how the phase transition gets affected in the presence of magnetic field in GB gravity background in higher spacetime dimensions. The effect of the magnetic field in this phase transition is different from that in metal/superconductor phase transition. For metal/superconductor phase transition, one gets critical magnetic field $B_c$ whereas in this case one gets a relation between a constant magnetic field $B$ and chemical potential $\mu$. \\  
\noindent In this paper, we have investigated the effect of magnetic field in presence of GB gravity for holographic insulator/superconductor phase transition in higher dimensional spacetime. We have carried out the investigation both numerically as well as analytically. To see the phase transition from insulator to superconductors, we consider the GB $AdS_{d}$ soliton background. We also consider the symmetric gauge to see effect of magnetic field on this phase transition in GB gravity background. First we employ two numerical approaches, namely, quasinormal mode method and the shooting method to study the critical phenomena. Both these approaches are based on the idea of marginally stable modes \cite{ssg},\cite{rlhy}. In both these techniques, one finds that marginally stable modes emerge at some  critical value of the chemical potential and the magnetic field. The emergence of such marginal stable modes indicate that the $AdS_{d}$ soliton background becomes unstable and a condensate of charged scalar field forms. Then we analytically investigate the same phenomena using St$\ddot{u}$rm-Liouville (SL) eigenvalue method and the analytical results agree with numerical results. It is observed that the square of the critical chemical potential and the magnetic field satisfies a linear relation $\Lambda^2\equiv\mu^2-B$. The value of $\Lambda^2$  increases with higher values of GB parameters $\tilde{\alpha}$ and dimension of spacetime $d$. This shows that phase transition becomes difficult in the presence of magnetic field for higher values of $\tilde{\alpha}$ and $d$.\\
\noindent The paper is organized as follows. In section 2, we discuss the basic set up of holographic insulator/superconductor phase transition in the presence of magnetic field. We investigate numerically critical phenomena using quasinormal mode in section 3. In section 4, we do same analysis using the shooting method. In section 5, we analytically investigate the critical phenomena in presence of magnetic field using the SL eigenvalue method. Finally, we conclude in section 6.

\section{Set up in the Gauss-Bonnet AdS$_{d}$ soliton background}
\noindent In this section, we construct the model of the holographic insulator to superconductor phase transition in the background of the Gauss-Bonnet $AdS_{d}$ soliton background. The metric for Gauss-Bonnet $AdS_d$ soliton reads \cite{Cai}
\begin{eqnarray}
ds^2_{d}=\frac{1}{f(r)}dr^2+r^2(-dt^2+ dx^2+ dy^2+dx_{i} dx^{i})+f(r)d{\chi}^2
\label{3}
\end{eqnarray}
with
\begin{eqnarray}
f(r)=\frac{r^2}{2\widetilde{\alpha}}\Bigg[1-\sqrt{1-\frac{4\widetilde{\alpha}}{L^2}\bigg(1-\frac{r_{0}^{d-1}}{r^{d-1}}\bigg)}\Bigg]
\label{4}
\end{eqnarray}
\noindent where $dx_{i}dx^{i}$, $(i=1,..,d-5)$, represents the line element of $(d-5)$-dimensional hypersurface with no curvature, $r_{0}$ is the tip of the soliton, $\widetilde{\alpha}$ is related to the GB coupling constant $\alpha$ as $\widetilde{\alpha}=(d-3)(d-4)\alpha $ and $L$ is the AdS radius. Without any horizon, this space time only have a conical singularity at $r=r_{0}$. Imposing a period $\beta=\frac{4\pi L^2}{(d-1) r_{0}}$ for the coordinate $\chi$ one can remove this singularity.\\
\noindent This line element is the solution of following action
\begin{eqnarray}
S=\frac{1}{16\pi G_{d}}\int d^{d}x \sqrt{-g}\Big[R-2\Lambda+\alpha R_{GB}\Big] 
\end{eqnarray}
where $R$ is the Ricci scalar, $R_{GB}=R^2-4R_{\mu\nu}R^{\mu\nu}+R_{\alpha \beta \gamma \delta }R^{\alpha \beta \gamma \delta }$ is the Gauss-Bonnet term and $\Lambda=-(d-1)(d-2)/(2L^2)$ is the cosmological constant.
The asymptotic behaviour of $f(r)~$ reads
\begin{eqnarray}
f(r)\sim\frac{r^2}{2\widetilde{\alpha}}\Bigg[1-\sqrt{1-\frac{4\widetilde{\alpha}}{L^2}}\Bigg]
\label{5}
\end{eqnarray}
with the effective asymptotic AdS scale defined by 
\begin{eqnarray}
L_{eff}^2=\frac{2\widetilde{\alpha}}{1-\sqrt{1-\frac{4\widetilde{\alpha}}{L^2}}} ~.
\label{6}
\end{eqnarray}
\noindent It should be noted that $L_{eff}^{2}= L^{2}$ and $L_{eff}^{2}= \frac{L^{2}}{2}$ for $\widetilde{\alpha}\rightarrow 0 $ and $\widetilde{\alpha}\rightarrow \frac{L^2}{4} $ respectively. The Schwarzschild AdS soliton is recovered by taking the limit $\widetilde{\alpha}\rightarrow 0$ in eq.(\ref{4}).\\
\noindent The matter Lagrangian for a holographic model of insulator/superconductor phase transition reads
\begin{eqnarray}
\mathcal{L}_{m} = -\frac{1}{4} F^{\mu \nu} F_{\mu \nu} - (D_{\mu}\psi)^{*} D^{\mu}\psi-m^2 \psi^{*}\psi
\end{eqnarray}
where $F_{\mu \nu}=\partial_{\mu}A_{\nu}-\partial_{\nu}A_{\mu}$ is the field strength tensor, $D_{\mu}\psi=\partial_{\mu}\psi-iqA_{\mu}\psi$ is the covariant derivative,  $A_{\mu}$ and $ \psi $ represent the gauge and the scalar fields.\\
\noindent The equations of motion of matter fields and gauge fields are 
\begin{eqnarray}
\frac{1}{\sqrt{-g}}D_{\mu}\Big[\sqrt{-g} g^{\mu \nu }D_{\nu}\psi\Big]-m^2\psi=0
\label{7}
\end{eqnarray}
\begin{eqnarray}
\frac{1}{\sqrt{-g}}\partial _{\mu}\Big\{\sqrt{-g}F^{\mu\nu}\Big\}=iq \Big[\overline{\psi}D^{\nu }\psi-\psi\overline{D^{\nu}\psi}\Big].
\label{8}
\end{eqnarray}
\noindent To solve these equations we need boundary conditions for these fields. From AdS/CFT correspondence, we know that the asymptotic behaviour of the fields are related to operators in the boundary theory in the following way \cite{ssg}
\begin{eqnarray}
\psi=\frac{\psi^{(-)}}{r^{\Delta_{-}}}+\frac{\psi^{(+)}}{r^{\Delta_{+}}}
\label{9}
\end{eqnarray}
\begin{eqnarray}
A_{t}=\mu-\frac{\rho}{r^{d-3}}
\label{10}
\end{eqnarray}
with 
\begin{eqnarray}
\Delta_{\pm}=\frac{1}{2}\left\{(d-1)\pm\sqrt{(d-1)^2+4m^{2}L_{eff}^{2}}\right\}
\end{eqnarray}
where $\Delta_{\pm}$ are the conformal dimensions,  $\mu$ and $\rho$ are interpreted as the chemical potential and charge density in the boundary field theory . In this work we consider $\psi^{(-)}=0$, so $\psi^{(+)}$ is related to the condensation operators in the boundary field theory.\\
\noindent To study the effect of the magnetic field in insulator/superconductor phase transition, we take the following ansatz
\begin{eqnarray}
A=\mu dt-\frac{B}{2}y dx+\frac{B}{2}x dy
\label{11}
\end{eqnarray}
which satisfies the gauge field equation (\ref{8}) and the boundary condition (\ref{10}).\\
\noindent  Introducing $z=\frac{r_{0}}{r}$ and considering an ansatz of the form $\psi=F(t,z)R(x,y)H(\chi)$, we obtain from eq.(\ref{7})
\begin{eqnarray}
\frac{d^2H(\chi)}{dt^2}=-\lambda ^2H(\chi)
\label{13}
\end{eqnarray}
\begin{eqnarray}
\frac{\partial ^{2}R}{\partial x^2}+\frac{\partial ^{2}R}{\partial y^2}-\frac{1}{4}q^2B^2(x^2+y^2)R+i q B\left(y\frac{\partial R}{\partial x}-x\frac{\partial R}{\partial y}\right )=-k^2R
\label{14}
\end{eqnarray}
\begin{eqnarray}
\frac{\partial ^{2}F}{\partial z^{2}}+\left(\frac{f^{\prime}(z)}{f(z)}-\frac{d-4}{z} \right )\frac{\partial F}{\partial z}-\frac{1}{z^2f}\frac{\partial ^2F}{\partial t^2}+\frac{2iq\mu }{z^2f}\frac{\partial F}{\partial t}\nonumber\\
+\frac{1}{z^2f}\left(q^2\mu^2-\frac{m^2r_{0}^2}{z^2}-\frac{\lambda ^2 r_{0}^2}{z^2f}-k^2 \right )F=0. 
\label{12}
\end{eqnarray}
\noindent where $\lambda=\frac{(d-1)l r_{0}}{2L^2}$, $l\in\mathbb{Z}$ to get periodicity of $H(\chi)=H\Big(\chi+\frac{4 \pi L^2}{(d-1)r_{0}}\Big)$. Since $\psi$ is axis symmetric, the last term in the left hand side of eq.(\ref{14}) is zero. Hence eq.(\ref{14}) becomes Schr$\ddot{o}$dinger like equation with two-dimensional harmonic potential, having eigenvalue $k^2=qB(n_{x}+n_{y}+1)$ where $n_{x},n_{y} \in \mathbb{Z}^+$. We expect that the lowest mode $l=0,n_{x}=0,n_{y}=0 $ will be the first most stable solution after condensation. Setting $L=1$ and $r_{0}=1 $, we obtain the equation of motion of $F(t,r)$ to be
\begin{eqnarray}
\frac{\partial ^{2}F}{\partial z^{2}}+\left(\frac{f^{\prime}(z)}{f(z)}-\frac{d-4}{z} \right )\frac{\partial F}{\partial z}-\frac{1}{z^2f}\frac{\partial ^2F}{\partial t^2}+\frac{2iq\mu }{z^2f}\frac{\partial F}{\partial t}+\frac{1}{z^2f}\left(q^2\mu^2-qB-\frac{m^2}{z^2} \right )F=0~. \nonumber \\
\label{15}
\end{eqnarray}
The solution of eq.(\ref{14}) reads \cite{rlhy}
\begin{eqnarray}
R(x,y)=e^{-\frac{q B}{4}(x^2+y^2)}
\label{15a}
\end{eqnarray}
which clearly shows the superconducting condensate will be localized to a finite circular region for any finite magnetic field. The region grows for smaller value of magnetic field and it occupies the whole $xy-$plane when $B\rightarrow 0$.  

\section{Critical behaviour via quasinormal modes}
\noindent In this section we study the critical behaviour via quasinormal modes in GB $AdS_{d}$ soliton background. The analysis of quasinormal modes of the perturbation in a fixed background provides a nice way of getting information about the stability of background spacetime. It turns out that the temporal part of the quasinomal modes behave like $e^{-i\omega t}$. Hence if the imaginary part of $\omega$ is negative, the mode decays in time. This means that the perturbation fades away thereby signalling the stability of the spacetime background. The reverse situation occurs when the imaginary part of $\omega$ is positive. The situation when $\omega=0$ is the critical case and the mode of the perturbation is said to be marginally stable. The existence of this mode is also expected to be a sign of instability \cite{ssg}. In \cite{cai1m},\cite{rlhy}, this method has been used to study critical behaviour in Einstein gravity. In this paper, we employ this method in the set up of GB gravity  to study the effect of magnetic field on holographic insulator/superconductor phase transition in arbitrary spacetime dimensions. 

\noindent In our analysis, we shall consider effects only up to first order in the GB parameter $\tilde{\alpha}$. Hence we expand the metric component $f(z)$ upto first order of GB parameter $(\tilde{\alpha})$ 
\begin{eqnarray}
f(z)=\frac{1-z^{d-1}}{z^2}\Big\{1+\tilde{\alpha }(1-z^{d-1})\Big\}+\mathcal{O}(\tilde{\alpha }^2) ~.
\label{15a}
\end{eqnarray}
In order to study the phase transition in this background, we further define
\begin{equation}
F(t,z)=e^{-i\omega t}W(z).
\end{equation}
Substituting this in eq({\ref{15}}), we get 
\begin{eqnarray}
\frac{d^{2}W}{d z^{2}}+\left(\frac{f^{\prime}(z)}{f(z)}-\frac{d-4}{z} \right )\frac{d W}{d z}+\frac{1}{z^4f(z)}\left(z^2(\omega+q \mu)^2-z^2 q B-m^2 \right )W=0 ~.
\label{16}
\end{eqnarray}
Multiplying throughout by $z^4 f(z)/(z-1)$ in the above equation, we obtain
\begin{eqnarray}
S(z)W^{\prime \prime}(z)+\frac{T(z)}{z-1}W^{\prime}(z)+\frac{V(z)}{(z-1)^2}W(z)=0
\label{17}
\end{eqnarray}
where the coefficients are given by
\begin{eqnarray}
S(z)&=&\frac{z^4f(z)}{z-1}\nonumber\\
    &=&-z^2(1+z+z^2+....+z^{d-2})\Big\{1+\tilde{\alpha }(1-z^{d-1})\Big\}\nonumber\\
T(z)&=&-(d-4)z^3f(z)+z^4f^{\prime}(z)\nonumber\\
    &=&-\Big[(d-2)z+z^{d}+\tilde{\alpha}\left(1-z^{d-1}\right)\Big\{(d-2)z+d z^{d}\Big\}\Big]\nonumber\\
V(z)&=&\big[z^2(q \mu+\omega)^2-z^2q B-m^2\big](z-1) ~.
\label{18}
\end{eqnarray}
 This coefficients $S(z),T(z),V(z)$ are all polynomials and can be written as 
\begin{eqnarray}
S(z)&=&\sum_{i=0}^{2d-1}s_{i}(z-1)^i\nonumber\\ T(z)&=&\sum_{i=0}^{2d-1}t_{i}(z-1)^i\nonumber\\
V(z)&=&\sum_{i=0}^{3}v_{i}(z-1)^i .
\label{19}
\end{eqnarray}
The coefficients $s_i,t_i,v_i$ can be calculated by comparing with eq.(\ref{18}). For $d=5$, the values of $s_{i},t_{i},v_{i}$ are given in Table \ref{tab0}. One can compute these values for other spacetime dimensions as well.
\begin{table}[h]
\caption{Non-zero $s_{i}$,$t_{i}$ and $v_{i}$ for marginal stable mode $(\omega=0)$  of  $d=5$, $(q=1)$}
\centering                          
\begin{tabular}{|c| c| c| c| c| c| c| c| c| c| c|}            
\hline
i&0&1&2&3&4&5&6&7&8&9\\
\hline
$s_{i}$&-4&-14+16$\tilde{\alpha}$&-20+80$\tilde{\alpha}$&-15+180$\tilde{\alpha}$&-6+240$\tilde{\alpha}$&-1+208$\tilde{\alpha}$&120$\tilde{\alpha}$&45$\tilde{\alpha}$&10$\tilde{\alpha}$&$\tilde{\alpha}$\\
\hline
$t_{i}$&-4&-8+32$\tilde{\alpha}$&-10+160$\tilde{\alpha}$&-10+400$\tilde{\alpha}$&-5+620$\tilde{\alpha}$&-1+628$\tilde{\alpha}$&420$\tilde{\alpha}$&180$\tilde{\alpha}$&45$\tilde{\alpha}$&5$\tilde{\alpha}$\\
\hline
$v_{i}$&0&$-m^{2}+\mu^{2}-B$&$2(\mu^2-B)$&$(\mu^2-B)$&0&0&0&0&0&0\\
\hline
\end{tabular}
\label{tab0}
\end{table} \\
\noindent Eq.(\ref{17}) is a second order differential equation with a regular singular point $z=1$. Hence one can write down a power series solution of this equation near the tip as
\begin{eqnarray}
W(z)=\lim_{N\rightarrow \infty }\sum_{j=0}^{N}a_{j}(z-1)^j ~.
\label{20}
\end{eqnarray}
\noindent Substituting eq.(s)(\ref{19}, \ref{20}) into  eq.(\ref{17}), we find  
\begin{eqnarray}
 \sum _{i=0}^{2d-1}s_{i}(z-1)^i\sum_{j=0}^{\infty}a_{j}j(j-1)(z-1)^{j-2}+\sum_{i=0}^{2d-1}t_{i}(z-1)^i\sum_{j=0}^{\infty}a_{j}j(z-1)^{j-2}\nonumber\\
+\sum _{i=0}^{3}v_{i}(z-1)^i\sum_{j=0}^{\infty}a_{j}(z-1)^{j-2}=0 .
\label{20a}
\end{eqnarray}
This in turn implies
\begin{eqnarray}
 \sum _{i=0}^{\infty }\sum _{j=0}^{\infty }a_{j}\Big\{ j(j-1)s_{i}+jt_{i}+v_{i}\Big\}(z-1)^{i+j}=0 ~.
\label{20b}
\end{eqnarray}
In the above equation, we have set the upper limit of $i$ to be $\infty$ using  the fact that $s_{j}$ , $t_{j}$ is zero when $j>(2d-1)$ and $v_{j}$ is zero if $j>3$. Introducing a new variable $n=i+j$, the above equation takes the form
\begin{eqnarray}
\label{20c}
\sum _{n=0}^{\infty }\sum _{j=0}^{n}a_{j}\Big\{ j(j-1)s_{n-j}+jt_{n-j}+v_{n-j}\Big\}(z-1)^{n}=0 \nonumber \\
\Rightarrow\sum_{j=0}^{n}a_{j}\Big\{j(j-1)s_{n-j}+jt_{n-j}+v_{n-j}\Big\}=0 ~.
\label{20d}
\end{eqnarray}
We now separate out the $n^{th}$-term from above equation to get a recursion relation which reads
\begin{eqnarray}
a_{n}=\frac{1}{(d-1)n^2}\sum_{j=0}^{n-1}\Big[j(j-1)s_{n-j}+j t_{n-j}+v_{n-j}\Big]a_{j} ~.
\label{21}
\end{eqnarray}
      

\noindent We now set $a_{0}=1$ for simplicity and use the boundary condition of the scalar field $\psi$ at $z=0$, which reads 
\begin{eqnarray}
W(0)=\lim_{N\rightarrow \infty}\sum_{n=0}^{N}a_{n}(-1)^n=0. 
\label{22}
\end{eqnarray}
The solution of this algebraic equation with the $a_n$'$s$ given by eq.(\ref{21}) gives the value of $\Lambda^2=\mu^2-B$ which determines the stability of the system. The smallest value of $\Lambda$ is the most marginal stable mode.

\noindent In the subsequent numerical calculations, we restrict $q=1$ and $N=600$. From numerical results it is observed that if a marginally stable mode arises, the square of the chemical potential $\mu^2$ and magnetic field $B$ satisfy a linear relation, whose slope is unity and intercept with $\mu^2$ axis gives the square of the critical chemical potential in the absence of the magnetic field. As the magnetic field increases, the critical chemical potential becomes higher. So in the presence of the magnetic field, transition from insulator to superconductor will be more difficult. 
We have shown the  first three lowest critical $\Lambda_{n}$'s ($n$ denotes the ``overtone number") in Table \ref{tab1} for Einstein gravity $(\tilde{\alpha}=0)$ which exactly match with the previous findings \cite{rlhy}. In Table \ref{tab2}, we have shown our main findings capturing the effects of the GB parameter $(\tilde{\alpha})$ for various mass of the scalar fields $(m^2)$ in different spacetime dimensions $(d)$. For $d=5$ and $m^2=-2$, the values of $\Lambda$ are $2.8145$ for Einstein gravity $(\tilde{\alpha}=0)$ and  $2.8493$ for GB gravity $(\tilde{\alpha}=0.02)$. The value of $\Lambda$ increases for higher values of GB parameter $(\tilde{\alpha})$. For $m^2=-2$ and $\tilde{\alpha}=0.01$, the values of $\Lambda$ are $2.832$ and $3.6378$ for $d=5$ and $d=6$ respectively which implies that the value of $\Lambda$ increases with increase in the number of spacetime dimensions $(d)$. This implies that the critical chemical potential increases for a fixed magnetic field.  This in turn shows that the phase transition becomes difficult in presence of the magnetic field in GB gravity background in higher spacetime dimensions. 
  
\begin{table}[h]
\caption{First three lowest $\Lambda_{n}$'s for Einstein gravity via quasinormal modes (QNM) and shooting method (SM).}   
\centering                          
\begin{tabular}{|c| c| c| c| c| c|}            
\hline                                 
$ d $ & $m^2$ & & \multicolumn{3}{c|}{$\Lambda_{n}$}  \\
\hhline{~~~---}
& & & $\Lambda_{0}$ & $\Lambda_{1}$ & $\Lambda_{2}$ \\
\hline
5 & 0.0 &QNM& 3.4041 & 5.8760& 8.3051\\ 
& & SM& 3.4037& 5.8763& 8.3057\\
\hhline{~-----}
  & -2 &QNM& 2.8145 & 5.2456& 7.6554\\ 
  & & SM& 2.8146& 5.2456& 7.6554\\ 
\hhline{~-----}
 & -15/4 &QNM& 1.8849 & 4.2263& 6.6032\\ 
 & & SM& 1.8884& 4.2345& 6.6160\\
\hline
6 & 0.0 &QNM& 4.0613& 6.6881& 9.2490\\ 
& & SM& 4.0612& 6.6878& 9.2491\\   
\hhline{~-----}
   & -2 &QNM& 3.6188 & 6.2107& 8.7538\\ 
   & & SM& 3.6188& 6.2107& 8.7538\\
\hhline{~-----}
 & -15/4 &QNM& 3.1325 & 5.6806& 8.2041\\ 
 & & SM& 3.1325& 5.6806& 8.2041\\ 
  \hline
  \end{tabular}
\label{tab1}  
\end{table}  

\begin{table}[h]
\caption{First three lowest $\Lambda_{n}$'s for Gauss-Bonnet gravity via the quasinormal modes (QNM) and shooting method (SM).}
\centering                       
\begin{tabular}{|c| c| c| c| c| c| c| c| c| c| c| c|}
\hline
$d$ & $m^2$ & & \multicolumn{3}{c|}{$\tilde{\alpha}=0.0001$} & \multicolumn{3}{c|}{$\tilde{\alpha}=0.01$} & \multicolumn{3}{c|}{$\tilde{\alpha}=0.02$} \\
\hhline{~~~---------}

  &  & &{$\Lambda_{0}$} & {$\Lambda_{1}$} & {$\Lambda_{2}$}& {$\Lambda_{0}$} & {$\Lambda_{1}$} & {$\Lambda_{2}$}& {$\Lambda_{0}$} & {$\Lambda_{1}$}& {$\Lambda_{2}$}\\
  \hline 
5 & 0.0 & QNM & 3.4042 & 5.8762 & 8.3054 & 3.417 & 5.8962 & 8.3332 & 3.4300 & 5.9163 & 8.3611 \\
 &  & SM & 3.4037 & 5.8763 & 8.3057& 3.4173 & 5.8962 & 8.3338 & 3.4295 & 5.9161 & 8.3619 \\
\hhline{~-----------}
    &-2 & QNM & 2.8147 & 5.2459 & 7.6557 & 2.832 & 5.2711 & 7.6889 & 2.8493 & 5.2964 & 7.7222 \\
    &  & SM & 2.8147 & 5.2459 & 7.6557 & 2.832 & 5.2711 & 7.6889 & 2.8493 & 5.2964 & 7.7222 \\
\hhline{~-----------}
    & -15/4 & QNM & 1.8854& 4.2269 & 6.6039 & 1.9291 & 4.2834 & 6.6699 & 1.9701 & 4.3367 & 6.7326 \\
    & & SM & 1.8888 & 4.2351 & 6.6167 & 1.9315& 4.2895& 6.6797 & 1.9719& 4.3413 & 6.7401 \\
    \hline
6  & 0.0& QNM & 4.0614 & 6.6883& 9.2493 & 4.0773 & 6.7126 & 9.2824 & 4.0933 & 6.7370 & 9.3156  \\
 & & SM & 4.0612 & 6.6878 & 9.2492 & 4.0771 & 6.7140 & 9.2808 & 4.093 & 6.7358 & 9.3123  \\
\hhline{~-----------} 
    & -2 & QMN & 3.6190 & 6.2109 & 8.7542 & 3.6378 & 6.2386 & 8.7908 & 3.6566 & 6.2664 & 8.8275 \\
   & & SM & 3.6190 & 6.2109 & 8.7542 & 3.6378 &6.2386  & 8.7908 & 3.6566 & 6.2664 & 8.8276\\
\hhline{~-----------}
    & -15/4 & QNM& 3.1327& 5.6809& 8.2045 & 3.1563 & 5.7143 & 8.2471 & 3.1799 & 5.7477 & 8.2896 \\
    & & SM& 3.1327 & 5.6809 & 8.2045 & 3.1563 & 5.7144 & 8.2471 & 3.1799 & 5.7477 & 8.2896 \\
    \hline
\end{tabular}
\label{tab2}
\end{table}

\section{Critical behaviour via shooting method}
\noindent An alternative way to numerically study the critical behaviour of the phase transition is the so called shooting method \cite{rlhy}. Here we describe how to use the shooting method to study the critical behaviour in GB $AdS_{d}$ soliton background. Using the shooting method we plot the profile of the scalar field and compare it with the results of quasinormal modes approach. 
We consider the static case in which $F$ is independent of $t$. Hence setting $\omega=0$ in eq.(\ref{16}), we obtain 
\begin{eqnarray}
W^{\prime\prime}(z)-\Bigg[\frac{dz^{d-1}-1}{z(1-z^{d-1})}+\frac{d-1}{z\Big\{1-\frac{\tilde{\alpha}}{1+\tilde{\alpha}}z^{d-1}\Big\}}\Bigg] W^{\prime}(z)-\frac{\frac{m^2}{z^2}-(q^2\mu^2-qB)}{(1-z^{d-1})\Big\{1+\tilde{\alpha }(1-z^{d-1})\Big\}}W(z)=0~. \nonumber \\
\label{23a}  
\end{eqnarray}
To study the behaviour of the solution near the tip of the soliton $z=1$, we substitute $f(z)\approx\frac{1-z^{d-1}}{z^2}$ in eq.(\ref{16}). This gives
\begin{eqnarray}
W^{\prime\prime}(z)-\frac{1}{1-z}W^{\prime}(z)+\frac{k}{(d-1)(1-z)}W(z)=0
\label{24}  
\end{eqnarray}
where $k\equiv q^2\mu^2-qB-m^2$. The solution of this equation near $z=1$ reads
\begin{eqnarray}
W(z)\mid _{z\rightarrow 1}=\alpha+\beta \log \left \{ \frac{k}{d-1}(1-z) \right \}
\label{25}  
\end{eqnarray}
where $\alpha$ and $\beta$ are two constants. Since the field is finite at the tip, we have to impose the condition $\beta=0$.

\noindent Near the boundary $W(z)$ behaves as
\begin{eqnarray}
W(z)\mid _{z\rightarrow 0}=\psi ^{(-)}z^{\frac{1}{2}\left\{(d-1)-\sqrt{(d-1)^2+4m^{2}}\right\}}+\psi ^{(+)}z^{\frac{1}{2}\left\{(d-1)+\sqrt{(d-1)^2+4m^{2}}\right\}} ~.
\label{26}  
\end{eqnarray}
In the following calculations, we will set $\psi ^{(-)}=0$ in order to turn off the effect of the source on the boundary field theory.

\noindent Setting $q=1$ for simplicity, we note that at the critical point of the phase transition, $W(z)$ is very close to zero. Therefore, we impose the following conditions at the tip $z=1$
\begin{eqnarray}
W(1)=0.001,~~ W^{\prime}(1)=\frac{k}{d-1}W(1) ~.
\label{27}  
\end{eqnarray}
The second condition follows from eq.(\ref{24}). For a given $d$, $m^2$ and $\tilde{\alpha}$, we now solve eq.(\ref{23a}) by the shooting method. We start with the above initial value of $W(z)$ at the tip $z=1$ and then numerically solve eq.(\ref{23a}), such that the condition $\psi ^{(-)}=0$ is satisfied at the boundary. This fixes the values of $\Lambda$. This $\Lambda$ implies that particular combinations of chemical potential $\mu$ and magnetic field $B$ satisfy the matter field equation for $\psi$. The values of $\Lambda^2\equiv\mu^2-B$ are shown in Table \ref{tab1} and Table \ref{tab2} for Einstein and GB gravity respectively. For $d=5$ and $m^2=-2$, the values $\Lambda$ are $2.8146$ for Einstein gravity $(\tilde{\alpha}=0)$ and $2.8493$ for GB gravity $(\tilde{\alpha}=0.02)$. Further, the values of $\Lambda$ increases for higher values of GB parameter  $(\tilde{\alpha})$. For $m^2=-2$ and $\tilde{\alpha}=0.01$, the values of $\Lambda$ are $2.832$ and $3.6378$ for $d=5$ and $d=6$ respectively. This implies that the value of $\Lambda$ increases for higher values of spacetime dimensions $(d)$ as well. The results are in very good agreement with those obtained from QNM method.
We also plot the scalar field profile for the lowest three $\Lambda_n$ for different values of $\tilde{\alpha}, d,$ and $m^2$ in Figure 1. 
From Table \ref{tab2} we find that $\Lambda_{0}$ increases with higher spacetime dimension, GB parameter and mass of the scalar field. This implies that condensation in GB gravity background is harder than in Einstein background, and it becomes more difficult as the number of spacetime dimensions increases and the mass of the scalar field increases.      
\begin{figure}[h!]
\caption{Scalar field profile for lowest three $\Lambda_n $ in different set of parameters.}
\includegraphics[scale=0.35]{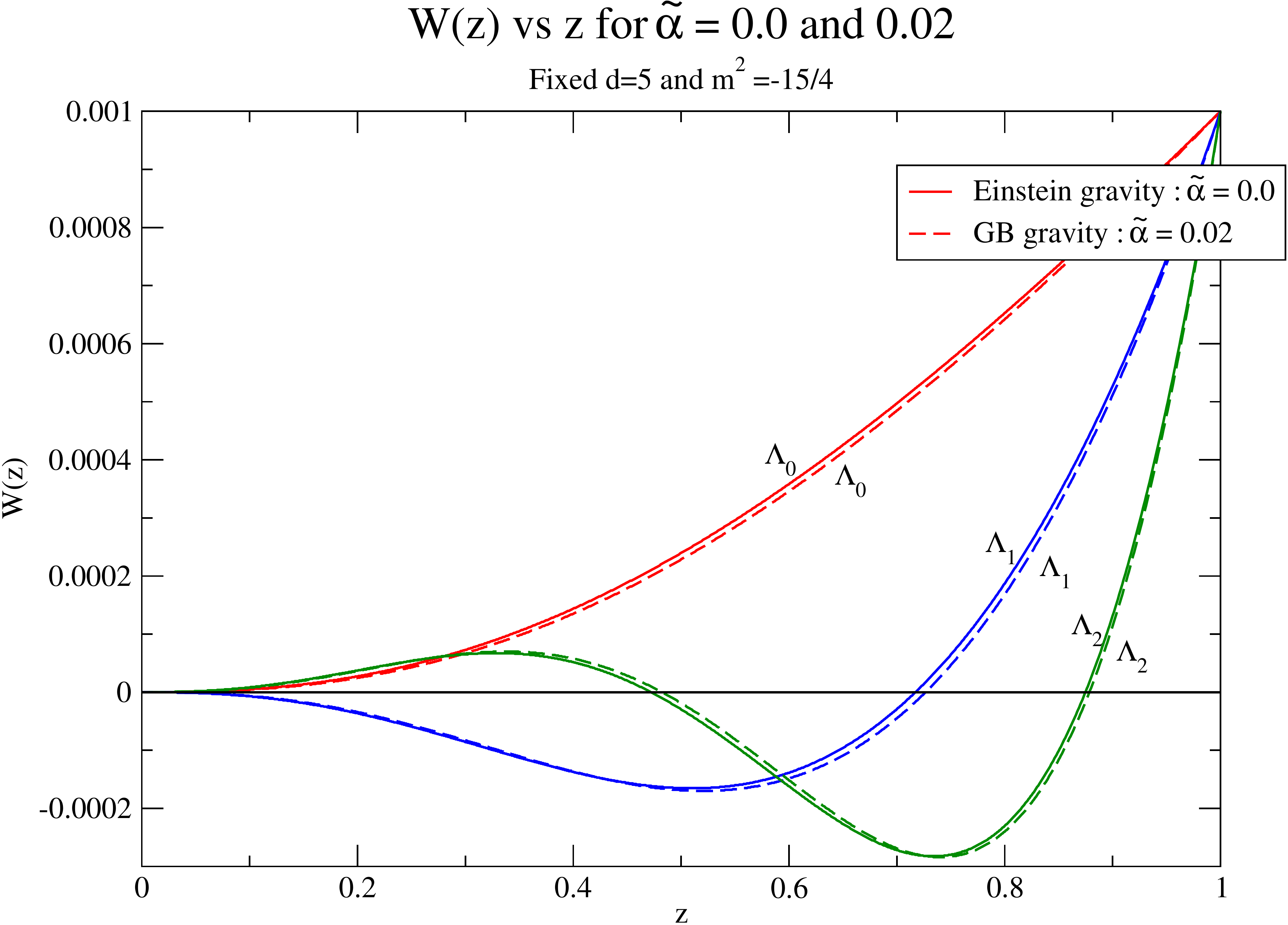} 
\includegraphics[scale=0.25]{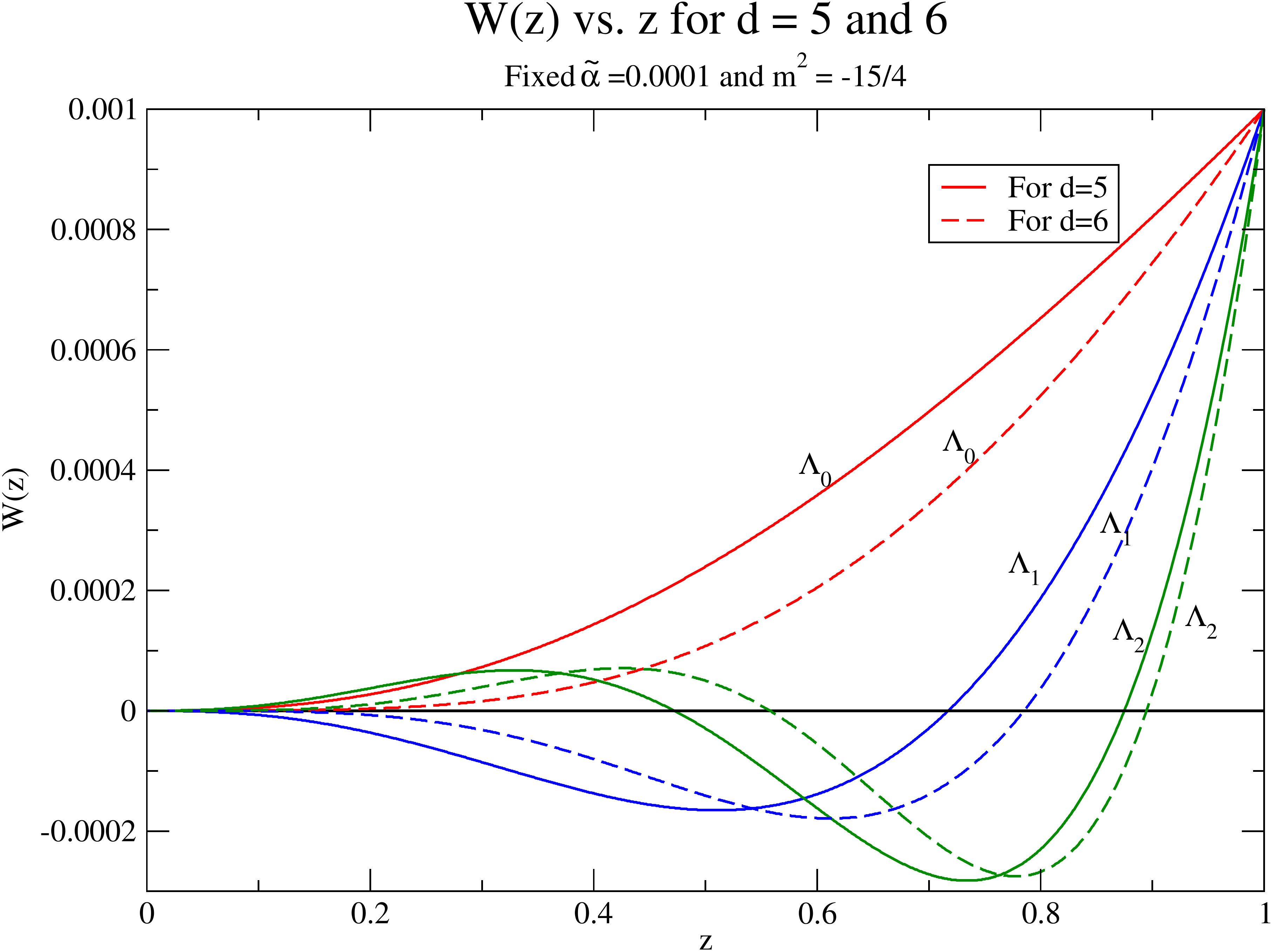} 
\includegraphics[scale=0.25]{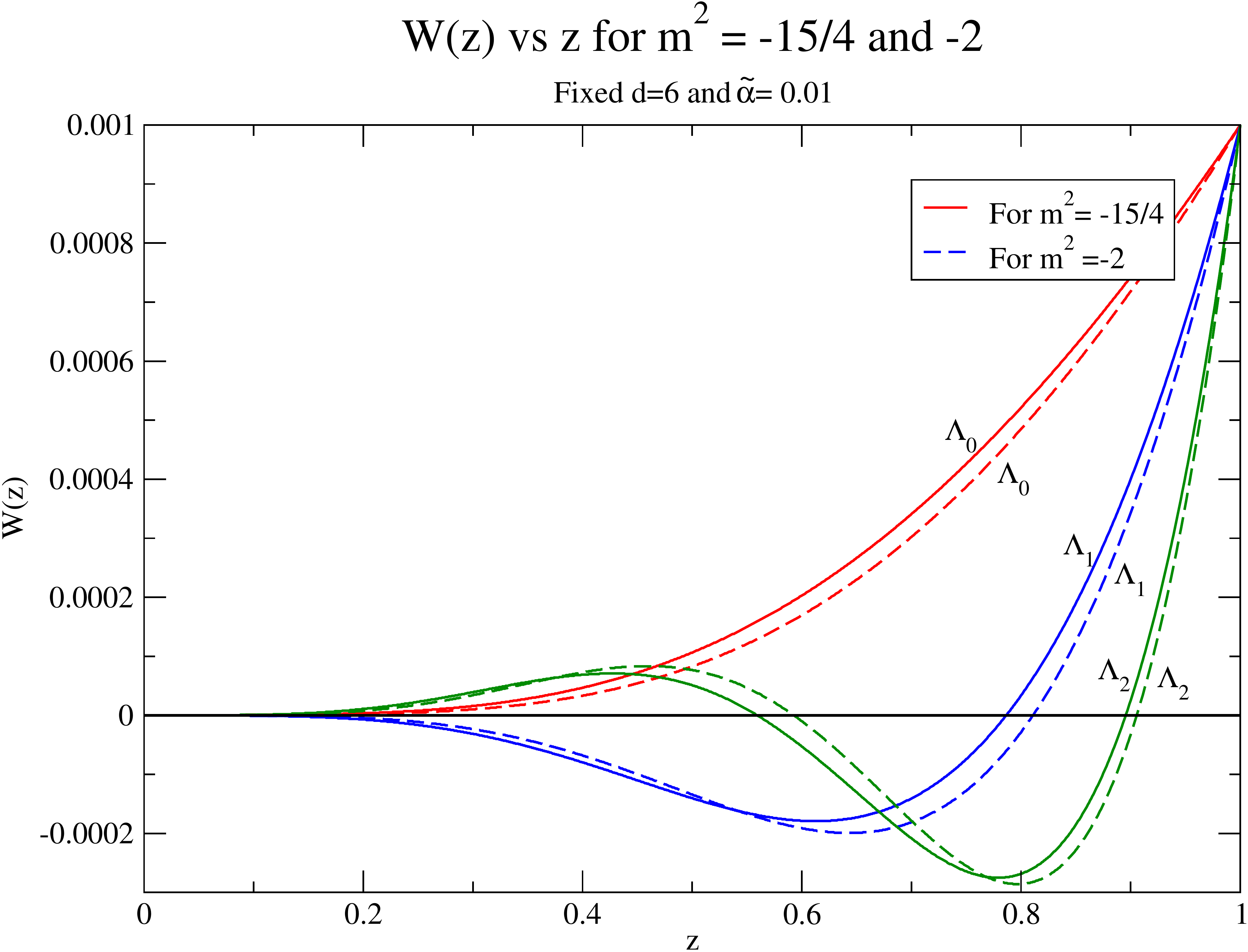}
\end{figure}

\section{Critical behaviour via the St$\ddot{u}$rm-Liouville method}

\noindent In  this  section  we  study analytically the critical  behaviour  using  St$\ddot{u}$rm-Liouville method.  From the last two sections, we observed that when the combination of chemical potential $\mu$ and magnetic field $B$, which is $\Lambda^2\equiv\mu^2-B$, exceeds a critical value $\Lambda_{0}$  for given mass, dimension and GB parameter, the condensations of the operators will happen. This can be viewed as a superconductor phase. For  $\Lambda<\Lambda_{0}$ ,the scalar field is zero and this can be interpreted as the insulator phase. Therefore, the critical parameters satisfying $\Lambda_{0}^2=\mu^2-B$, are the turning points of the holographic insulator/superconductor phase transition. Here we are trying to find an approximate function to relate the parameters ${q,\mu,B,m^2,d,\widetilde{\alpha}}$ near the critical phase transition point. Starting from eq(\ref{23a}), we introduce a trial function $\Gamma(z)$ into $W(z)$ near $z=0$ as
\begin{eqnarray}
W(z)\sim\langle{\mathcal{O}_{\pm}\rangle}z^{\Delta_{\pm}}\Gamma(z)
\label{28}  
\end{eqnarray}
 satisfying $\Gamma(0)=1$. Substituting eq.(\ref{28}) in eq.(\ref{23a}), we obtain 
\begin{eqnarray}
\Gamma^{\prime\prime}(z)+\Bigg\{\frac{2\Delta_{\pm}}{z}+\left( \frac{f^{\prime}(z)}{f(z)}-\frac{d-4}{z}\right)\Bigg\}\Gamma^{\prime}(z)+\Bigg\{\frac{\Delta_{\pm}(\Delta_{\pm}-1)}{z^2}\nonumber\\
+\frac{\Delta_{\pm}}{z}\bigg(\frac{f^{\prime}(z)}{f(z)}-\frac{d-4}{z}\bigg)+ \frac{1}{z^{4}f(z)} \left((q^2\mu^2-qB)z^{2}-m^{2}\right)\Bigg\}\Gamma(z)=0
\label{29}
\end{eqnarray}
to be solved subject to the boundary condition $\Gamma^{\prime}(0)=0$.
\noindent The above equation can be recast in the St$\ddot{u}$rm-Liouville form
\begin{eqnarray}
\frac{d}{dz}\big\{p(z)\Gamma^{\prime}(z)\big\}+q(z)\Gamma(z)+(q^2\mu^2-qB)r(z)\Gamma(z)=0
\label{30}
\end{eqnarray}
with
\begin{eqnarray}
&p(z)&=\frac{z^{2\Delta_{\pm}-d+2}}{2\widetilde{\alpha}}\bigg[1-\sqrt{1-4\widetilde{\alpha}\left(1-z^{d-1}\right)}\bigg]\nonumber\\
&q(z)&=p(z)\Bigg[\frac{\Delta_{\pm}(\Delta_{\pm}-1)}{z^2}+\frac{\Delta_{\pm}}{z}\Bigg\{ \frac{(d-5) z^{d-2}-2 \big[z f(z)-\frac{2}{z}\big]}{2 \left(z^{d-1}-1\right)+z^2f(z)}-\frac{d-4}{z}\Bigg\}-\frac{m^2}{z^{4}f(z)}\Bigg]\nonumber\\
&r(z)&=z^{2\Delta_{\pm}-d+2} ~.
\label{31}
\end{eqnarray}
To estimate the minimum eigenvalue $q^2\mu^2-qB$, we write down an equation for $q^2\mu^2-qB$, extremization of which leads to eq.(\ref{30}).
 This reads
 \begin{eqnarray}
q^2\mu^2-qB=\frac{\int_{0}^{1}dz\big\{p(z)[\Gamma^{\prime}(z)]^{2}-q(z)[\Gamma(z)]^{2}\big\}}{\int_{0}^{1}dz r(z)[\Gamma(z)]^{2}} ~.
\label{32}
\end{eqnarray}
We now use the trial function 
\begin{eqnarray}
\Gamma(z)=1-a z^2
\label{33}
\end{eqnarray}
which satisfies the boundary conditions $F(0)=1$ and $F^{\prime}(0)=0$, where $a$ is a constant. Substituting eq(s).(\ref{33}) and (\ref{31}) in eq.(\ref{32}), we get
\begin{eqnarray}
q^2\mu^2-qB=\frac{U(\widetilde{\alpha},d,m)-V(\widetilde{\alpha},d,m)a+S(\widetilde{\alpha},d,m)a^2}{\frac{1}{2\Delta-d+3}-\frac{2a}{2\Delta-d+5}+\frac{a^2}{2\Delta-d+7}}
\label{34}
\end{eqnarray}
where
\begin{eqnarray}
U(\widetilde{\alpha},d,m)&=&\frac{1}{2\Delta-d+1}\Bigg[m^2+\frac{\Delta^2}{2\widetilde{\alpha}}\bigg\{1-\left(\sqrt{1-4\widetilde{\alpha}}\right)\, _2F_1\left(-\frac{1}{2},\frac{2\Delta-d+1}{d-1};\frac{2\Delta}{d-1};\frac{4 \widetilde{\alpha}}{4 \widetilde{\alpha}-1}\right)\bigg\}\Bigg]\nonumber\\
V(\widetilde{\alpha},d,m)&=&\frac{2}{2\Delta-d+3}\Bigg[m^{2}+\frac{\Delta ^2}{2\widetilde{\alpha}}\bigg\{1-\left(\sqrt{1-4 \widetilde{\alpha}}\right) \, _2F_1\left(-\frac{1}{2},\frac{2\Delta-d+3}{d-1};\frac{2 (\Delta +1)}{d-1};\frac{4\widetilde{\alpha}}{4\widetilde{\alpha}-1}\right)\bigg\}\nonumber\\
&-&\frac{\Delta (2\Delta-d+1)}{2\widetilde{\alpha}}\bigg\{1-\left(\sqrt{1-4 \widetilde{\alpha}}\right) \, _2F_1\left(\frac{1}{2},\frac{2\Delta-d+3}{d-1};\frac{2 (\Delta +1)}{d-1};\frac{4\widetilde{\alpha}}{4\widetilde{\alpha}-1}\right)\nonumber\\
&-&\frac{(4\Delta-d+1)(2\Delta-d+3)}{(\Delta+1)(2\Delta-d+1)}\left(\frac{\widetilde{\alpha}}{\sqrt{1-4 \widetilde{\alpha}}}\right)\, _2F_1\left(\frac{1}{2},\frac{2 (\Delta +1)}{d-1};\frac{2\Delta +d+1}{d-1};\frac{4\widetilde{\alpha}}{4\widetilde{\alpha}-1}\right)\bigg\}\Bigg]\nonumber\\
S(\widetilde{\alpha},d,m)&=&\frac{1}{2\Delta-d+5}\Bigg[m^{2}+ \frac{\Delta ^{2}+4}{2\widetilde{\alpha}}-\frac{2\sqrt{1-4 \widetilde{\alpha}}}{\widetilde{\alpha}}\left(1+\frac{\Delta ^2}{4} \right ) \, _2F_1\left(-\frac{1}{2},\frac{2\Delta-d+5}{d-1};\frac{2 (\Delta +2)}{d-1};\frac{4\widetilde{\alpha}}{4\widetilde{\alpha}-1}\right)\nonumber\\
&-&\frac{\Delta (2\Delta-d+1)}{2\widetilde{\alpha}}\bigg\{1-\left(\sqrt{1-4 \widetilde{\alpha}} \right)\, _2F_1\left(\frac{1}{2},\frac{2\Delta-d+5}{d-1};\frac{2 (\Delta +2)}{d-1};\frac{4\widetilde{\alpha}}{4\widetilde{\alpha}-1}\right)\nonumber\\
&-&\frac{(2\Delta-d+5)(4\Delta-d+1)}{(\Delta+2)(2\Delta-d+1)}\frac{\widetilde{\alpha}}{\sqrt{1-4 \widetilde{\alpha}}}\, _2F_1\left(\frac{1}{2},\frac{2 (\Delta +2)}{d-1};\frac{2\Delta +d+3}{d-1};\frac{4\widetilde{\alpha}}{4\widetilde{\alpha}-1}\right)\bigg\}\Bigg] ~.
\label{35}
\end{eqnarray}
To estimate the value of the critical parameter $\Lambda_{0}^{2}\equiv(q^2\mu^2-q B)$, we need to minimize $(q^2\mu^2-q B)$ with respect to $a$. For $d=6$, $m^2=-2$ and $\tilde{\alpha}=0.01$, we obtain from eq.(\ref{34})
\begin{eqnarray}
q^2\mu^2-qB=\frac{18.0755 a^2-38.1719 a+23.2873}{ a^2-2.49183 a+1.65222}~.
\label{36}
\end{eqnarray}
$\Lambda_{0}^2$ is determined by minimizing $q^2\mu^2-qB$ with respect to $a$. It is found that $\Lambda_0$ attains minimum value at $a=0.53$ and $\Lambda_{0}=3.6443$. Similarly, we have calculated other values of $\Lambda_{0}^2$ for different $m$,~$d$ and $\tilde{\alpha}$.
In Table \ref{tab3}, we show the values of $\Lambda_0$ (lowest one) for various values of $m^2$, $d$ and GB parameter which match with the numerical findings. 
In the absence of the magnetic field, we find the critical chemical potential which agrees with previous findings \cite{dp2}. For Einstein gravity we also recover previous findings in \cite{rlhy}.  For $m^2=-2$ and $\tilde{\alpha}=0.01$, the values of $\Lambda_0$ are $2.8348$, $3.6443$ and $4.3759$ for $d=5,6,7$ respectively which implies that the value of $\Lambda_0$ increases with increase in the number of spacetime dimensions $(d)$. It is observed that the critical parameter $\Lambda_{0}$ increases with larger value of spacetime dimension, mass of the scalar field and GB parameter. Our findings obtained through the SL approach are consistent with the results obtained by the other approaches discussed in this paper.
\begin{table}[h!]
\caption{Analytical values of critical parameters $\Lambda_{0}$ for different values of $\tilde{\alpha}, d$ and $m^2$ }   
\centering                          
\begin{tabular}{|c| c| c| c| c|}            
\hline                                 
$ d $ & $m^2$ & \multicolumn{3}{c|}{$\tilde{\alpha}$}  \\
\hhline{~~---}
& & 0.0001 & 0.01 & 0.02 \\
\hline
 & 0 & 3.4069 & 3.42& 3.4336  \\ 
\hhline{~----}
5 & -2 & 2.8173 & 2.8348 & 2.8530  \\ 
\hhline{~----}
 & -$\frac{15}{4}$ & 1.8908 & 1.9344& 1.9771 \\ 
\hline
 & 0 & 4.068 & 4.0842 & 4.1010   \\    
\hhline{~----}
6 & -2 & 3.6252 & 3.6443 & 3.6641  \\
\hhline{~----}
 & -$\frac{15}{4}$ & 3.1383 & 3.1623 & 3.1873  \\ 
  \hline
& 0 & 4.7134 & 4.7326 & 4.7526   \\    
\hhline{~----}
7 & -2 & 4.3545 & 4.3759 & 4.3981  \\
\hhline{~----}
 & -$\frac{15}{4}$ & 3.9944 & 4.0187 & 4.0439  \\
\hline   
\end{tabular}
\label{tab3}  
\end{table}
\section{Conclusions}
\noindent We now summarize our findings. In this paper, we have made an investigation on the effect of the magnetic field on holographic insulator/superconductor phase transition in higher dimensional spacetime in $AdS_{d}$ soliton background. The importance of this investigation is to see the effect of higher curvature correction and magnetic field in this phase transition. 
Based on the idea of the marginal stable mode method, the quasinormal mode method, the shooting method and the analytical St$\ddot{u}$rm-Liouville method are adopted to study the critical phenomena. We observe that marginal modes emerge at critical values of the chemical potential and the magnetic field satisfying the relation $\Lambda^2 =\mu^2-B $. The values of $\Lambda$ increase with higher values of the GB parameter and dimension of the spacetime. In the absence of magnetic field, the critical chemical potential agrees with our previous finding \cite{dp2}. In the presence of magnetic field, we find that the condensation becomes harder because we find that the value of the chemical potential increases (since the value of $\Lambda$ increases) with increase in the values of the GB parameter, number of spacetime dimensions and mass of the scalar field. The analytical results agree very well with the numerical results. 

\section*{Acknowledgments}
DP would like to thank CSIR for financial support. DG would like to thank DST-INSPIRE for financial support. SG acknowledges the support under the Visiting Associateship programme of IUCAA, Pune.


\begin{thebibliography}{99}
\baselineskip=0.6 cm
\bibitem{adscft1} J. M. Maldacena, Adv. Theor. Math. Phys. 2 (1998) 231.
\bibitem{adscft2} E. Witten, Adv. Theor. Math. Phys. 2  (1998) 253.
\bibitem{adscft3} S.S. Gubser, I.R. Klebanov, A.M. Polyakov, Phys. Lett. B 428 (1998) 105.
\bibitem{ssg} S.S. Gubser, Phys. Rev. D 78 (2008) 065034.
\bibitem{adscft5} S.A. Hartnoll, C.P. Herzog, G.T. Horowitz, Phys. Rev. Lett. 101 (2008) 031601.
\bibitem{hs9} S. A. Hartnoll, C. P. Herzog, G. T. Horowitz, JHEP 12 (2008) 015.
\bibitem{hs8} R. Gregory, S. Kanno, J. Soda, JHEP 0910 (2009) 010.
\bibitem{siop} G. Siopsis, J. Therrien, JHEP 05 (2010) 013.
\bibitem{hs17} G. T. Horowitz, M. M. Roberts, JHEP  0911 (2009) 015.
\bibitem{hs9a} H.B. Zeng, X. Gao, Y. Jiang, H.-S. Zong, JHEP 05 (2011) 002.
\bibitem{hs9b} H.F. Li, R.-G. Cai, H.-Q. Zhang, JHEP 04 (2011) 028.
\bibitem{hs14}  Q. Y. Pan, B. Wang, E. Papantonopoulos, J. Oliveira, A. Pavan, Phys. Rev. D 81  (2010) 106007.
\bibitem{hs15}  R.G. Cai, H. Zhang, Phys. Rev. D 81 (2010) 066003.
\bibitem{horowitz} G. T. Horowitz and B. Way, J. High Energy Phys. 11
(2010) 011.
\bibitem{hs19} J. Jing, S. Chen, Phys. Lett. B 686 (2010) 68.
\bibitem{hs20} J. Jing, Q Pan, S. Chen, JHEP 1111 (2011) 045. 
\bibitem{sgdc1} S.~Gangopadhyay, D.~Roychowdhury, JHEP 05 (2012) 002.
\bibitem{hs24} S.~Gangopadhyay, D.~Roychowdhury, JHEP 05 (2012) 156.
\bibitem{sg-single}S.~Gangopadhyay, Phys. Lett. B 724 (2013) 176.
\bibitem{rb} R. Banerjee, S.~Gangopadhyay, D.~Roychowdhury, A. Lala, Phys. Rev. D 87 (2013) 104001.
\bibitem{sgm} S. Gangopadhyay, Mod. Phys. Lett. A 29 (2014) 1450088.
\bibitem{dg3} D.~Ghorai, S.~Gangopadhyay, Eur. Phys. J. C 76 (2016) 702.
\bibitem{dg1} D. Ghorai, S. Gangopadhyay, Eur. Phys. J. C 76 (2016) 146. 
\bibitem{dg4} D.~Ghorai, S.~Gangopadhyay,  Euro. Phys. Lett. 118 (2017) 31001.
\bibitem{dg6} D.~Ghorai, S.~Gangopadhyay, Nucl. Phys. B933 (2018) 1-13.
\bibitem{Hart}S. A. Hartnoll, Classical Quantum Gravity 26 (2009) 224002.
\bibitem{nrt} T. Nishioka, S. Ryu, and T. Takayanagi, J. High
Energy Phys. 03 (2010) 131.


\bibitem{cai-so} R.G. Cai, H.F. Li, H.Q. Zhang, Phys. Rev. D 83 (2011) 126007.
\bibitem{qjb} Qiyuan Pan, Jiliang Jing, and Bin Wang,  J. High
Energy Phys. 11 (2011) 088.
\bibitem{lee} C. Oh Lee, Eur. Phys. J. C 72 (2012) 2092.
\bibitem{cai1m} R.G. Cai, Xi He, H-F Li H.Q. Zhang, Phys. Rev. D 84 (2011) 046001.
\bibitem{rlhy} R.G. Cai, Li Li, H.Q. Zhang, Y.L. Zhang, Phys. Rev. D 84 (2011) 126008.
\bibitem{dp2} D. Parai, S. Gangopadhyay, D. Ghorai, Annals Phys. 403 (2019) 59.
\bibitem{Cai} R.G. Cai, S.P. Kim, and B. Wang, Phys. Rev. D 76 (2007) 024011.
\end{thebibliography}
\end{document}